\documentclass[aps,pre,twocolumn,groupedaddress,superscriptaddress,showpacs,longbibliography]{revtex4-1}
\pdfoutput=1

\usepackage{graphicx}
\usepackage{color}
\usepackage{amsmath}
\usepackage{amsfonts}
\usepackage{amssymb}
\usepackage{amsmath}
\usepackage{dcolumn}
\usepackage{hyperref}
\hypersetup{colorlinks=true,urlcolor=blue,linkcolor=blue,citecolor=blue}

\usepackage{mathptmx}
\usepackage[T1]{fontenc}

\begin{document}
\title{How to share underground reservoirs}

\author{K. J. Schrenk}
	\email{jschrenk@ethz.ch}
	\affiliation{Computational Physics for Engineering Materials, IfB, ETH Zurich, Wolfgang-Pauli-Strasse 27, CH-8093 Zurich, Switzerland}
    
\author{N. A. M. Ara\'ujo}
	\email{nuno@ethz.ch}
	\affiliation{Computational Physics for Engineering Materials, IfB, ETH Zurich, Wolfgang-Pauli-Strasse 27, CH-8093 Zurich, Switzerland}

\author{H. J. Herrmann}
	\email{hans@ifb.baug.ethz.ch}
	\affiliation{Computational Physics for Engineering Materials, IfB, ETH Zurich, Wolfgang-Pauli-Strasse 27, CH-8093 Zurich, Switzerland}
	\affiliation{Departamento de F\'isica, Universidade Federal do Cear\'a, Campus do Pici, 60451-970 Fortaleza, Cear\'a, Brazil}
\begin{abstract}
Many resources, such as oil, gas, or water, are extracted from porous soils and their exploration is often shared among different companies or nations.
We show that the effective shares can be obtained by invading the porous medium simultaneously with various fluids.
Partitioning a volume in two parts requires one division surface while the simultaneous boundary between three parts consists of lines.
We identify and characterize these lines, showing that they form a fractal set consisting of a single thread spanning the medium and a surrounding cloud of loops.
While the spanning thread has fractal dimension ${1.55\pm0.03}$, the set of all lines has dimension ${1.69\pm0.02}$.
The size distribution of the loops follows a power law and the evolution of the set of lines exhibits a tricritical point described by a crossover with a negative dimension at criticality.
\end{abstract}
\maketitle
\section{\label{sec::intro}Introduction}
Space partitioning is of interest in a wide spectrum of fields, ranging from materials science to medicine, with special relevance to computer graphics and the exploration of natural resources stored in soils.
For example, if different companies want to explore an oil reservoir they are interested in determining the volumetric share corresponding to each one inside the ground \cite{Hannesson00}.
An additional degree of complexity comes into play when water is injected into the soil to push the oil to enhance extraction \cite{Christensen01, Maugeri09}.
Also in medical imaging, three-dimensional computed tomography scans need to be segmented to identify the different tissues.
These pictures are discretized into pixels and a number is assigned to the bond between neighboring pixels corresponding to the intensity gradient.
The resulting structure is similar to the one of a porous soil.
By aggregating pixels pairwise from the lowest to the highest gradient it becomes possible to identify the boundaries of tissues \cite{Yan06}.

\begin{figure}[b]
	\center\includegraphics[width=\columnwidth]{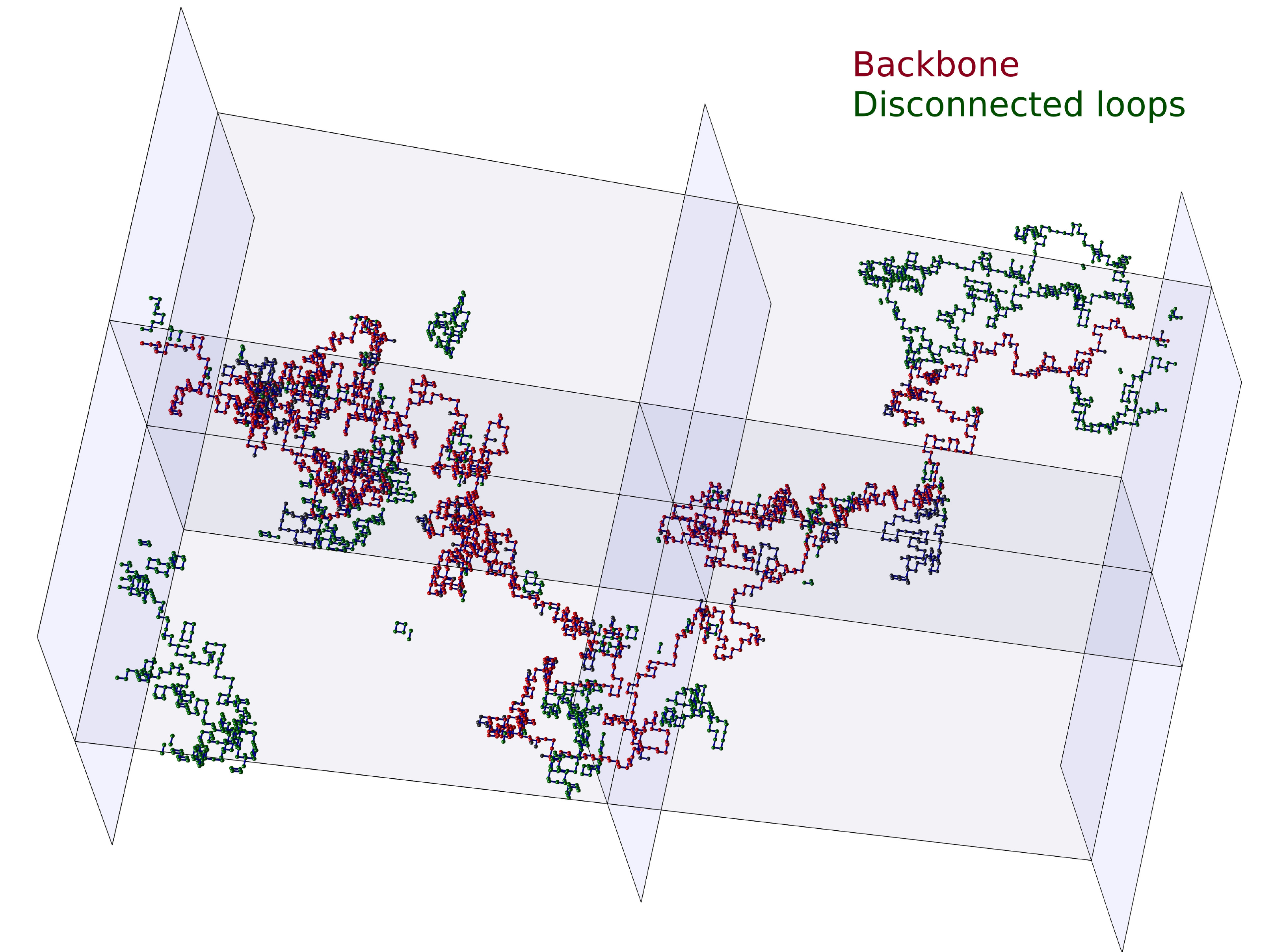} 
	\caption{
	Set of lines on which all division surfaces between three parts are in contact for a typical random medium.
	In addition to the backbone spanning the medium from left to right (shown in red), the set also contains a cloud of disconnected loops (green).
       	The transparent planes are guides to the eye.
       	\label{fig::Snap3d}
	}
\end{figure}

Both problems consist in dividing space into parts: either the shares of the companies in the oil field or the different tissues in the image processing.
In both cases, regions are separated by {\it division surfaces}.
Here we consider three regions and find that their division surfaces join in a fractal thread that crosses the medium, being surrounded by a cloud of disconnected loops (see Fig.~\ref{fig::Snap3d}).
In the case of oil exploration these points, where all three division surfaces merge, are the places where water should be injected to assure that no oil is pushed out on the wrong side.
In medical image processing, the simultaneous boundary between three parts might indicate, for example, the region where a tumor is nested between two other tissues.
\section{Model}
\subsection{Random media}
We consider a random medium consisting of pores arranged in a simple-cubic lattice connected through channels.
To each channel $k$ a threshold $p_k$ is randomly assigned following a uniform distribution in the interval $[0,1)$.
The fraction of open channels is tuned by a parameter $p$, such that channels with $p_k<p$ are open while all the others are closed.
Hereafter we use the language of fluids where $p$ would correspond to the fluid pressure.
For digital images, the pores would correspond to the pixels and the thresholds $p_k$ to the intensity gradient between pairs of neighboring pixels.
\subsection{Partitioning}
\begin{figure}
	\center\includegraphics[width=\columnwidth]{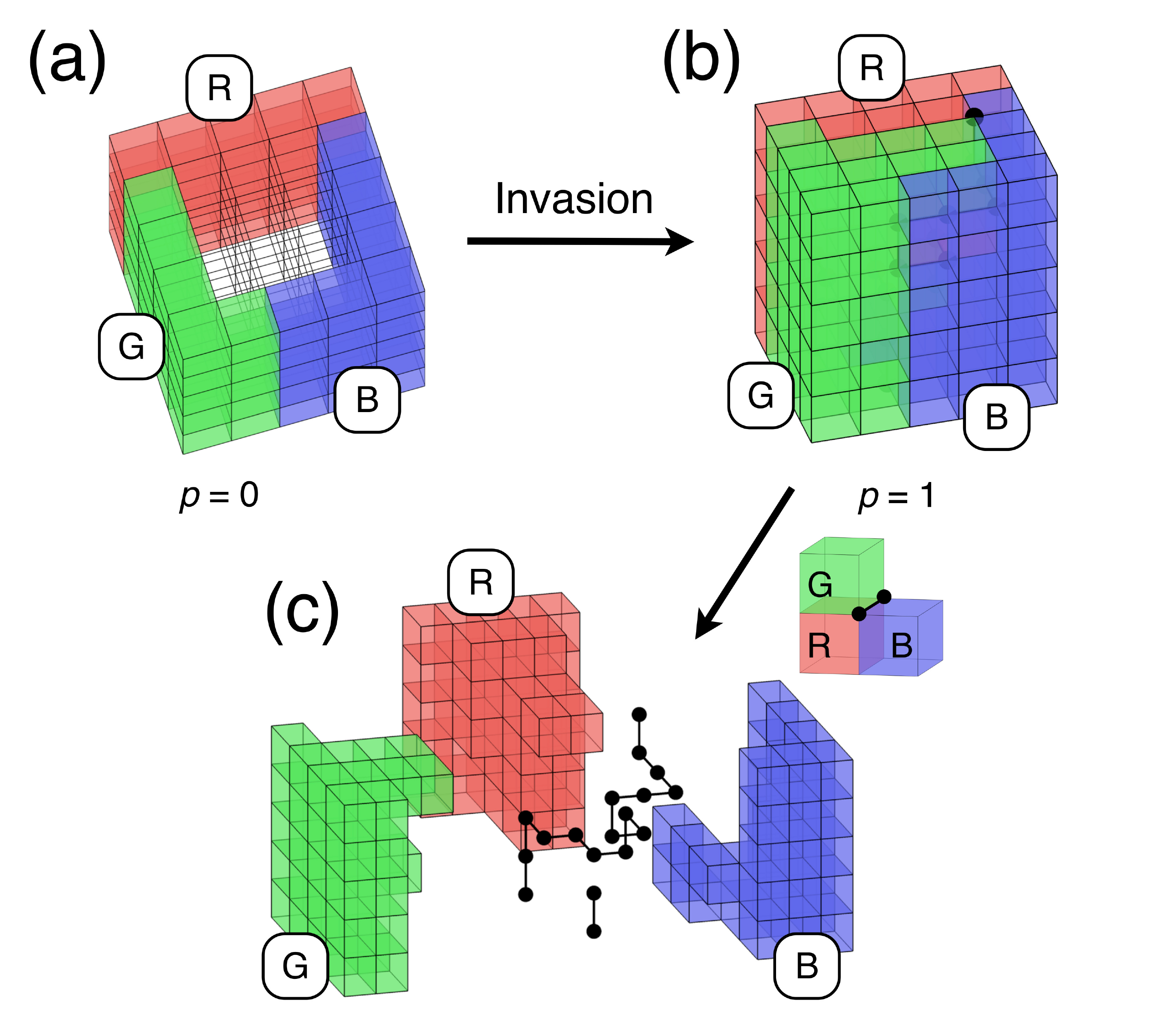}
	\caption{
        Illustration of the model.
        (a) In the initial state $(p=0)$ the vertical faces of the cubic lattice are divided into three sets.
        (b) Example of the final state of the invasion $(p=1)$, dividing the medium into three parts: R, G, and B.
        RGB edges and nodes are shown as thick black lines and spheres, respectively.
	In (c) we separate the three pieces to be able to look inside.
	Solid lines represent the edges of the dual lattice of the pore network.
	The color of each cube corresponds to the one of the fluid in the pore at the center of the cube.
	The channels connecting the pores are perpendicular to the faces of the cubes and for clarity they are not shown.
	The RGB edges and nodes are part of the dual lattice.
        \label{fig::LatticeDefinition}
        }
\end{figure}

To find the partitioning of the medium into three parts, we divide the (four) vertical boundaries of a cubic system in three parts of about the same area.
Each part corresponds to a different invading fluid distinguished by dyeing them with different colors: red (R), green (G), and blue (B) [see Fig.~\ref{fig::LatticeDefinition}(a)].
In the illustration of Fig.~\ref{fig::LatticeDefinition} we see a medium of $5\times5\times5$ pores.
The pores are in the center of each cube and the edges are the bonds of the dual lattice of the pore network.
The cubes have the color of the fluid contained in the corresponding pore.
We invade the system simultaneously from all vertical walls.

Starting with $p=0$ (i.e., all channels closed), the channel with the lowest threshold, in the invasion front, is selected and the fluid pressure $p$ is increased to the value of this threshold.
This channel and the empty pore connected to it are then invaded and colored according to the type of fluid that penetrated into it.
After that, invasion also cascades into all pores connected to this pore through channels with thresholds lower than the actual $p$.
This process is repeated until all pores are invaded, under the constraint that the fluids can not displace each other, which does not allow to invade any pore by more than one fluid.

In the final state, the medium is divided into three parts corresponding to the pores filled either with an R, G, or B fluid.
These parts are the maximum oil shares that each company could extract from the exploration regions.
This division is solely determined by the distribution of local thresholds, thus being intrinsic to the medium (for algorithmic details see Appendix \ref{sec::SimDet}).
Here we will mainly focus on the final partitioning of the medium, corresponding to $p=1$ (all pores invaded).
However, we later also discuss the pressure $p_t$ at which two different fluids start to form an interface.
\subsection{RGB set}
An example for the partitioning into three parts is shown in Fig.~\ref{fig::LatticeDefinition}(b).
To better visualize the partitioning, we separate the three parts in Fig.~\ref{fig::LatticeDefinition}(c).
Every face that separates two colors is part of a division surface.
Edges are shared by four different cubes.
If three of the cubes sharing a common edge have different colors, we call this an {\it RGB edge} (thick black lines in Fig.~\ref{fig::LatticeDefinition}).
In fact, all RGB edges are on lines where all three surfaces separating regions of different color meet.
The vertices attached to RGB edges are the {\it RGB nodes} and we define every set of nodes connected through RGB edges as an {\it RGB cluster}.
All RGB nodes and edges of a medium form its {\it RGB set}.
\section{Results}
\begin{figure}
        \center\includegraphics[width=\columnwidth]{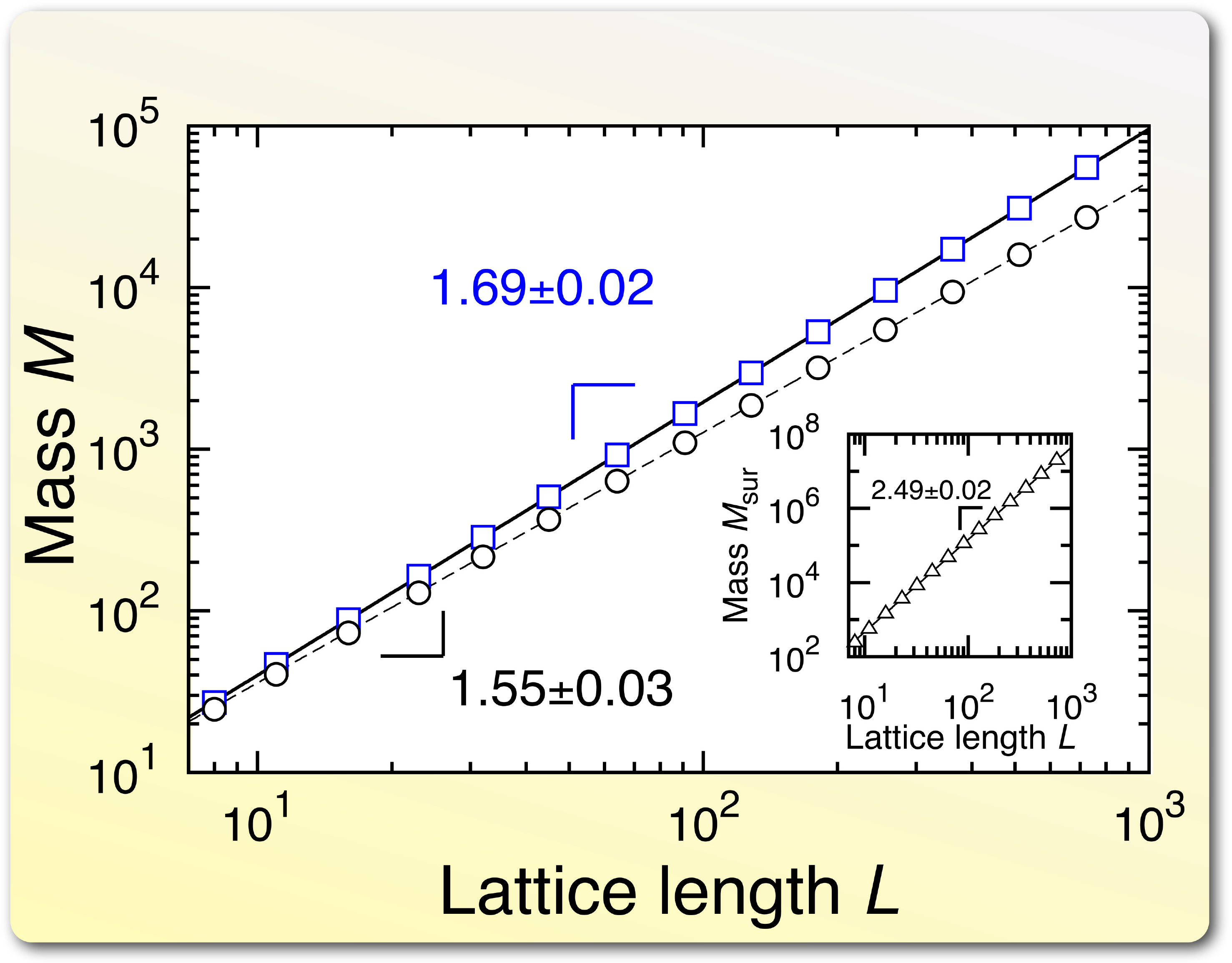}
        \caption{
        Mass scaling analysis.
        The total mass $M_\text{tot}$ ($\square$, total number of RGB nodes) and the mass of the spanning cluster $M_\text{sc}$ ($\bigcirc$, number of nodes in largest RGB cluster) are shown as function of the lattice length $L$.
        One observes that for large lattices the masses scale as power laws of the lattice size, i.e., $M_\text{tot}\sim L^{d_\text{tot}}$ and $M_\text{sc}\sim L^{d_\text{sc}}$.
        The fractal dimensions are $d_\text{tot}=1.69\pm0.02$ and $d_\text{sc}=1.55\pm0.03$.
        In the inset we see the area of the division surfaces $M_\text{sur}$ $(\triangle)$ as function of the lattice size $L$.
        The fractal dimension of the division surfaces is $d_\text{sur}=2.49\pm0.02$.
        Straight lines are guides to the eye.
	\label{fig::DataOnly}
	}
\end{figure}
The surfaces dividing two colors are fractal objects of fractal dimension $d_\text{sur}=2.49\pm0.02$, as seen in the inset of Fig.~\ref{fig::DataOnly}, numerically consistent with what has been reported for watersheds in three dimensions \cite{Cieplak94, Fehr11c}.
While these boundaries are singly connected, the RGB set consists of one spanning cluster connecting the two sides of the system surrounded by a cloud of smaller disconnected loops (see Fig.~\ref{fig::Snap3d}).
As shown in Fig.~\ref{fig::DataOnly}, the entire RGB set is fractal of dimension $d_\text{tot}=1.69\pm0.02$, while the spanning cluster has a smaller fractal dimension $d_\text{sc}=1.55\pm0.03$.
To analyze the topology of the spanning cluster, we used the burning method proposed in Ref.~\cite{Herrmann84}.
We found that the spanning cluster has loops, however its backbone, elastic backbone, shortest path, and its set of singly connected RGB edges all have fractal dimensions consistent with $d_\text{sc}$.

\begin{figure}
	\center\includegraphics[width=\columnwidth]{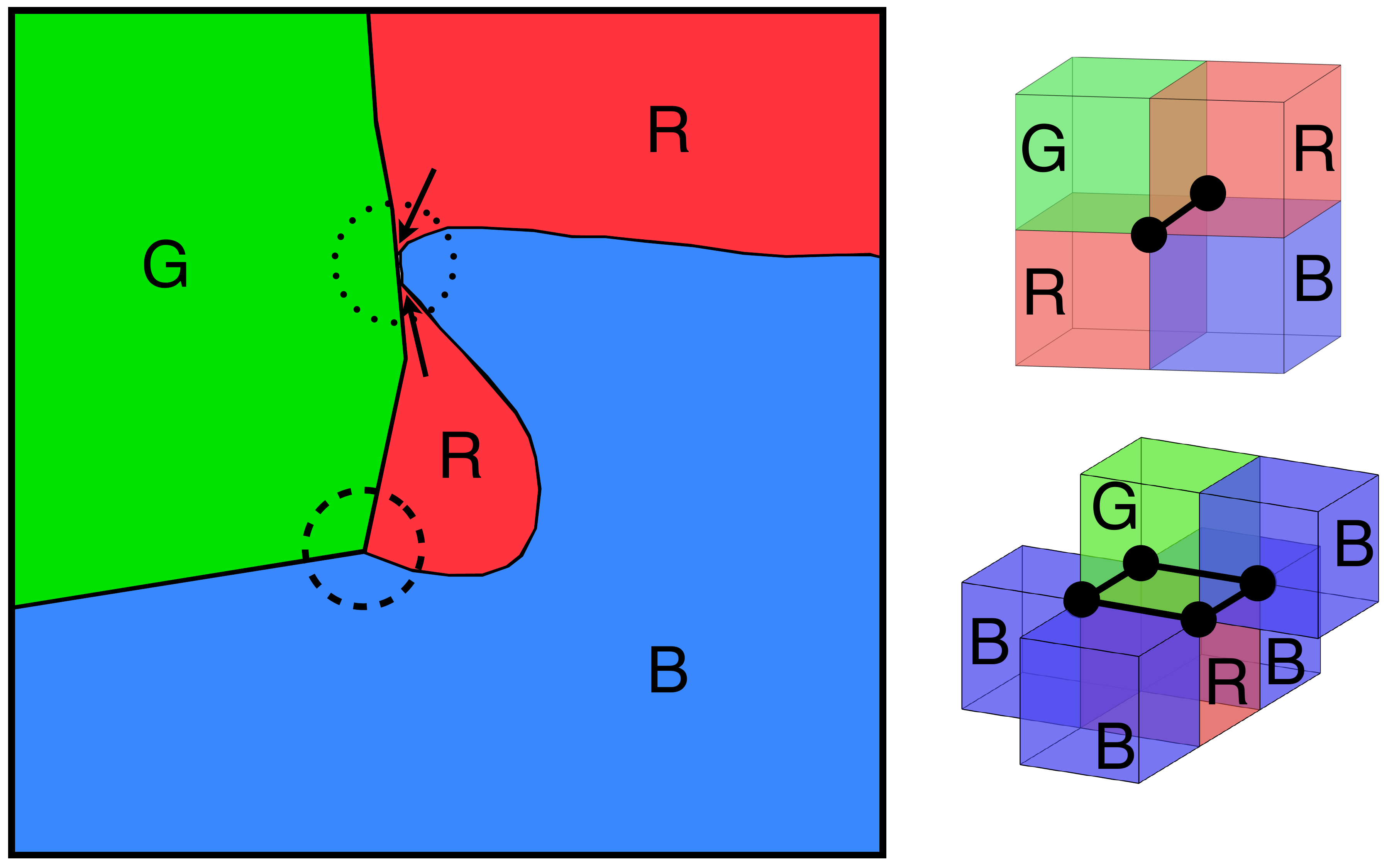}
        \caption{
        Sketch explaining the presence of disconnected loops in the RGB set.
        We see a horizontal cut through a partitioned medium.
        The two red regions are connected with each other somewhere above or below the shown plane.
        The arrows indicate the existence of a loop, as discussed in the main text.
        On the right, one sees examples of cubes of different colors sharing edges.
        RGB edges are shown as thick solid lines.
        \label{fig::SketchLoopsNoDangling}
        }
\end{figure}

\begin{figure*}
	\includegraphics[width=2\columnwidth]{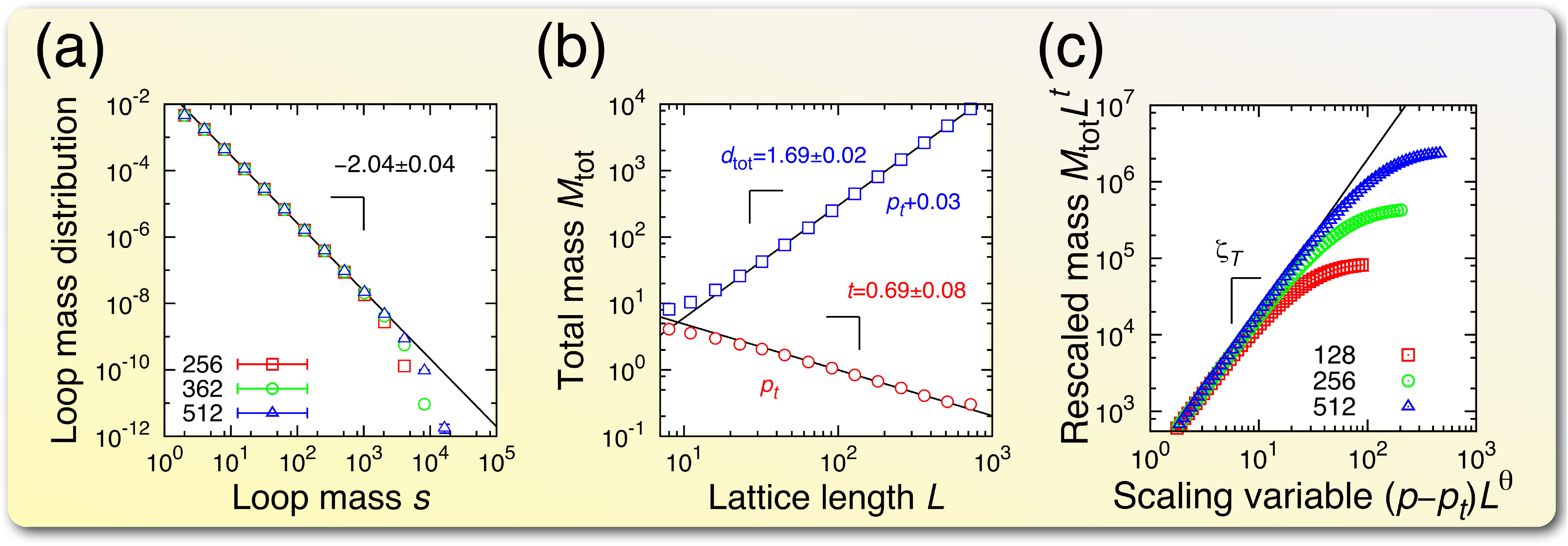}
    	\caption{
    	Scale-free behavior.
    	(a) Number of loop sizes $s$ divided by $L^{d_\text{tot}}$ as function of $s$ for different lattice sizes $L$ ($256$ $(\square)$, $362$ $(\bigcirc)$, and $512$ $(\triangle)$).
    	For intermediate sizes the data follows a power law, $p(s)\sim s^{-a}$ with $a=2.04\pm0.04$.
    	(b) Size dependence of the total mass $M_\text{tot}$ at and above the threshold pressure $p_t$.
    	At $p_t$ $(\bigcirc)$ the total mass decreases according to a power law, $M_\text{tot}\sim L^{-t}$, where $t=0.69\pm0.08$.
    	In contrast, above the threshold $(\square)$ the slope is given by the fractal dimension $d_\text{tot}$.
    	(c) Rescaled total mass $M_\text{tot}L^{t}$ as a function of the scaling variable $(p-p_t)L^\theta$ for different lattice sizes $L$.
    	Close to the threshold $M_\text{tot}\sim(p-p_t)^{\zeta_T}$, where $\zeta_T=2.0\pm0.3$.
    	The solid lines are guides to the eye.
    	\label{fig::band}
    	}
\end{figure*}

The difference between $d_\text{tot}$ and $d_\text{sc}$ is due to the cloud of disconnected loops.
These loops result from the entanglement of three compact regions, as illustrated in Fig.~\ref{fig::SketchLoopsNoDangling}, which shows a transversal cross section of a medium, where the three regions are simultaneously in contact at different locations.
In this particular case, the lower location (dashed circle) is where the spanning cluster intersects the cross section.
The upper location (dotted circle) shows the cut through a disconnected loop:
The G and B regions are in contact in an area completely surrounded by the R region, thus the contact line between the three forms a closed loop (discretization effects are discussed in the Appendix).
The size distribution of the loops is shown in Fig.~\ref{fig::band}(a), where the size $s$ is defined as the number of RGB nodes forming the loop.
A power-law distribution is observed, $p(s) \sim s^{-a}$, with $a=2.04\pm0.04$, revealing the absence of a characteristic size.
The distribution of distances of disconnected loops from the spanning cluster decays exponentially, i.e., the loop cloud is mainly localized in the neighborhood of the spanning cluster (see Appendix for data).

To understand how the RGB set emerges, we now consider its evolution with the control parameter $p$.
Initially, when the fluids R, G, and B start to invade from the boundary, the RGB set is empty.
As $p$ increases, at a typical value $p=p_t$, two fluids for the first time try to invade the same channel in the bulk and with increasing $p$ a division surface starts to form orthogonal to these channels.
If, in addition, any of the four edges shared by two neighboring pores of different color is also shared by a pore of the third color, an RGB edge emerges.
Figure \ref{fig::band}(b) shows how the total number of RGB nodes, $M_\text{tot}$, depends on system size at $p=p_t$ and at $p=p_t+0.03$.
While above $p_t$ results are consistent with the fractal dimension observed for the final state $(p=1)$, precisely at $p_t$, a negative scaling exponent is obtained, $M_\text{tot} \sim L^{-t}$, with $t=0.68\pm0.08$.
This implies that, in the thermodynamic limit, the RGB set is empty at $p_t$, while above $p_t$ it is fractal of fractal dimension $d_\text{tot}$.

If we assume that the formation of RGB nodes at $p_t$ is the product of two uncorrelated processes, namely the formation of a dividing surface between two colors, we can show that $t=d-2b$, where $d$ is the spatial dimension and $b$ is the fractal dimension of the dividing surfaces at $p_t$ (see Appendix).
For the fractal dimension $b$ of the set connecting two fluids at $p_t$, Coniglio has shown in the context of percolation that $b=1/\nu$, where $\nu$ is the correlation-length critical exponent of percolation \cite{Coniglio89}.
In three dimensions, $\nu=0.8734\pm0.0006$ \cite{Lorenz98, Ballesteros99, Deng05b}, giving $t=3-2/\nu\approx0.71$, consistent with our numerical result.
Above the threshold $p_t$ the argument leading to the expression for $t$ does not hold, since in this regime the invasion is correlated.
Accordingly, the fractal dimension of the RGB set is then different from the one of the intersection of the two division surfaces.
We conjecture that the expression for $t$ can be generalized to any dimension $d$ and number of different fluids (colors) $n$, as far as $2 \leq n \leq d$:
\begin{equation}
t(d,n) = (n-2)d-(n-1)/\nu.
\end{equation}
For two colors, $n=2$, $t=-1/\nu=-b$, as in percolation.
In contrast for $n\geq3$, given the exact and numerical values for $\nu$ \cite{Stauffer94}, $t$ is positive.
Above $d=6$, the upper-critical dimension of percolation, $1/\nu=2$, such that $t=(n-2)d-2(n-1)$.

We find that $M_\text{tot}$ scales with the distance to $p_t$ as ${M_\text{tot}\sim(p-p_t)^{\zeta_T}}$ with $\zeta_T=2.0\pm0.3$.
Therefore, we propose the following crossover scaling for the total number of RGB nodes:
\begin{equation}
\label{eq::scaling}
M_\text{tot}(p,L)=L^{-t}G[(p-p_t)L^\theta].
\end{equation}
This scaling behavior of $M_\text{tot}(p,L)$ in $p$ and $L$ implies ${\theta=(d_\text{tot}+t)/{\zeta_T}}$.
In addition, the scaling function $G[x]$ fulfills $G[x] \sim x^{\zeta_T}$ for $x>0$.
The {\it Ansatz} in Eq.~(\ref{eq::scaling}) is confirmed by the scaling plot in Fig.~\ref{fig::band}(c).
\section{Discussion}
We found a rich scale-free behavior in the partitioning of random media through simultaneous invasion by three fluids.
The lines where all three fluids are simultaneously in contact form a fractal set, the RGB set, of dimension $d_\text{tot}=1.69\pm0.02$, while its spanning cluster has dimension $d_\text{sc}=1.55\pm0.03$.
The other clusters are loops and their size follows a power-law distribution.
At the threshold where two fluids first form an interface, the size of the set of RGB nodes scales with a negative exponent in the system size.
We propose a crossover scaling between this exponent and the fractal dimension $d_\text{tot}$ above the threshold.

For an oil reservoir shared by three companies, our study shows how the optimal injection regions scale with the reservoir size and how they are spatially distributed.
In the second example, image analysis, our work establishes how the number of pixels forming the simultaneous boundary between three tissues scales with the image resolution.
Besides these examples, our results have also implications to other fields.
Let us consider a chemical reaction that requires three different reactants each entering a porous medium from another side.
Supposing that all reactants have the same diffusion constant our results identify the disconnected fractal region in which the reaction will take place.
Finally, knowing the properties of the partitioning of a porous medium is also relevant for planning of waste disposal.
Often trash is buried under ground, in porous soils, such that its decomposition gases spread through pores and fractures \cite{Wettstein12}.
These gases will eventually leave the soil and the partitioning of the soil determines where this will happen first.
In particular, the fractal RGB set are the disposal regions where the escaping of gases occurs, simultaneously, in three regions.

The exact shape of the RGB set depends on the threshold distribution and on the injection areas of the fluids.
We also tested the partition model with different sets of injection pores, namely,
(1) division of the six faces of a cubic medium into three injection areas, corresponding each to two adjacent faces of the cube,
(2) injection from three vertical faces of a cube, and
(3) injection from three edges of the cube, with periodic boundary conditions.
For all cases, we obtained fractal dimensions consistent with $d_\text{tot}=1.69\pm0.02$.
In contrast, for the following injection patterns, different values for $d_\text{tot}$ have been obtained:
(1) division of the six faces such that each fluid is injected from two opposite faces,
(2) injection pores uniformly distributed in the cube, with periodic boundary conditions, and
(3) three single injection pores in the cube, also with periodic boundaries.
These observations suggest that two conditions on the injection areas, though not necessary, are sufficient to obtain the RGB fractal dimension reported here.
First, the injection area of each color must be singly connected.
Second, the division of the surface of the medium into these areas has to be such that no single fluid can isolate the remaining two fluids from each other.

The reported fractal dimensions were obtained for a uniform and uncorrelated distribution of thresholds.
It is well-known that disorder in soils is typically characterized by spatial correlations, which can be described by their Hurst exponent $H$.
The numerical values of the fractal dimensions reported here will in general depend on $H$ \cite{Isichenko92, Sahimi93, Oliveira11, Fehr11b}.
Nevertheless, our qualitative and topological arguments should still be applicable.

Models of discontinuous percolation transitions are a subject of recent interest \cite{Achlioptas09, Riordan11, Manna10, Chen11, Nagler11, Nagler12, Bizhani12, Araujo10, Schrenk11, Schrenk11b}.
Some of these models lead to compact clusters with fractal perimeters \cite{Araujo10, Schrenk11, Schrenk11b} and in some cases with a fractal dimension compatible with the one of division surfaces \cite{Schrenk11}.
The simultaneous boundaries between three clusters are therefore quite likely related to RGB sets.
\appendix
\section{\label{sec::SimDet}Simulation details}
All numerical results have been obtained with Monte Carlo simulations.
Random numbers have been generated with the algorithm proposed in Ref.~\cite{Ziff97}.
Considering the labeling scheme by Newman and Ziff \cite{Newman00, Newman01}, we kept track of the color properties as function of the fraction of sampled channels $p$.
For Fig.~\ref{fig::DataOnly}, results have been averaged over at least $2800$ realizations.
In Fig.~\ref{fig::band}(a), (b), and (c), results have been averaged over at least $10000$, $6400$, and $300$ realizations, respectively.
Unless indicated otherwise, statistical error bars are smaller than the symbol size.
The algorithmic procedure shares similarities with invasion percolation \cite{Wilkinson83, Lenormand89} and the fracturing of ranked surfaces \cite{Schrenk12}.
The self-similarity of the shortest path in the spanning cluster of the RGB set has been confirmed using the yardstick method \cite{Mandelbrot83, Tricot88}.
\section{Scaling exponent $t$ and relation to random percolation}
In the context of the scaling behavior at the threshold $p_t$, in the Section {\it Results}, we give the following expression for the scaling exponent,
\begin{equation}
t=d-2b.
\end{equation}
It describes the scaling of the total number of RGB nodes with the lattice size as ${M_\text{tot} \sim L^{-t}}$.
This expression can be obtained under the assumption that the RGB network, which corresponds to the region where the division surfaces merge, results (only at $p_t$) from the intersection of two uncorrelated dividing surfaces.
Suppose that the division surfaces at $p_t$ are fractals of dimension $b$, i.e., ${M_\text{sur} \sim L^{b}}$ where $M_\text{sur}$ is the mass of the division surface (see explanation below).
Then, we can make use of a general result for the intersection of two fractal objects reported in Ref.~\cite{Miyazima87}.
Consider the intersection of two independent fractals of dimensions $d_{f,1}$ and $d_{f,2}$ in spatial dimension $d$, with their centers separated less than the bigger radius of gyration of the two. 
Then the fractal dimension of the intersection of both fractals is
\begin{equation}
	\label{eq::df_inter_12}
	d_{f,1\cap2}=d_{f,1}+d_{f,2}-d.
\end{equation}
Therefore, if one considers the intersection of two objects of dimension $b$, one has ${d_{f,1\cap2}=2b-d}$.
And, since $t$ is defined by ${M_\text{tot} \sim L^{-t}}$, 
\begin{equation}
	\label{eq::t_of_b}
	t=-d_{f,1\cap2}=d-2b.
\end{equation}

To find expressions for the fractal dimension of the division surfaces at the threshold and the value $p_t$ of the threshold itself, we apply the relation of our model to random percolation \cite{Stauffer94}.
If there would not be the constraint that the fluids cannot displace each other, percolation connecting to three sides \cite{Kleban06} would be recovered.
Then, a spanning cluster of invaded pores would emerge at a fraction of open channels ${p=p_c}$, where $p_c$ is the bond percolation threshold of the lattice.
Therefore, for $p$ increasing from zero, the first contact of two fluids in the bulk occurs at ${p=p_t=p_c}$, i.e., the contact pressure threshold in our model equals the bond percolation threshold of the lattice.
In this work, we have used the bond percolation threshold of the simple-cubic lattice, ${p_c=0.2488126}$ \cite{Lorenz98, Dammer04}.
Let us now consider the fractal dimension $b$ of the division surfaces at the threshold.
Below $p_t$, the three fluids are not in contact and the set of RGB nodes is empty.
Therefore, there exists no correlation between the emerging division surfaces between the fluids.
As a result, the RGB network behaves like the intersection between two independent division surfaces.
On the other hand, in a non-spanning configuration of critical percolation, bonds that, once occupied, would yield a spanning cluster are called anti-red bonds and their number diverges with the lattice size as $L^{1/\nu}$ \cite{Coniglio89,Scholder09}, where $\nu$ is the correlation-length critical exponent.
Now, opening any of the channels contained in the division surfaces of our model at ${p_t=p_c}$ would give rise to a cluster spanning the entire medium.
Since, in addition, these surfaces are independent, they have the fractal dimension of the anti-red bonds, i.e., ${b=1/\nu}$.
For our model, Eq.~(\ref{eq::t_of_b}) yields ${t=3-2/\nu\approx0.71}$.
By applying Eq.~(\ref{eq::df_inter_12}), the generalization of the expression for $t$  given in Eq.~(1) is obtained.
Equation (1) applies to our model for ${d=n=3}$.
There, since the sum of the fractal dimension of two division surfaces is smaller than the spatial dimension, the objects are mutually transparent (no intersections) and the number of bonds in such surfaces vanishes in the thermodynamic limit.
Above the threshold since correlations develop between neighboring surfaces, the considerations used to derive Eq.~(1) for $t$ do not hold.
In addition, from the numerical values of the fractal dimension, one then sees that the RGB network scales differently from the mere intersection of uncorrelated division surfaces.
\section{Loops and discretization effects}
In the article, we discuss that, for ${p=1}$, RGB clusters not connected to the largest one are loops.
An example how they emerge is shown in Fig.~4.
To understand possible discretization effects on the loops, expected for lattice studies, we analyze this example in more detail.
\begin{figure}[!h]
	\center
	\includegraphics[width=\columnwidth]{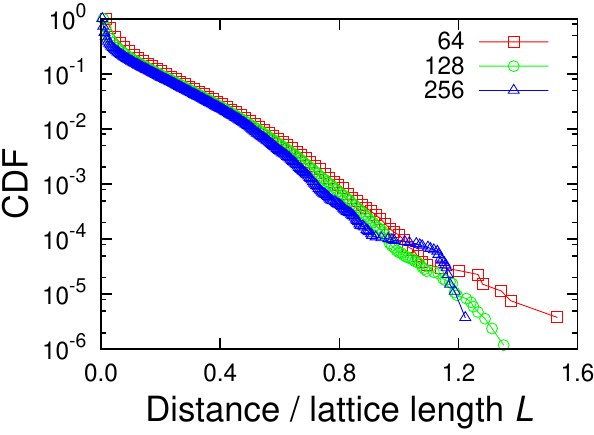}
	\caption{
		CDF of distances of disconnected loops from the spanning cluster.
		The number of loops with distance greater equal $k$ is plotted as function of ${k/L}$, for different lattice lengths $L$ ($64$ ($\square$), $128$ ($\bigcirc$), and $256$ ($\triangle$)). 
		\label{fig::LoopDistanceDistribution}
	}
\end{figure}
Figure 4 shows a sketch of a horizontal cross section through a medium.
Since ${p=1}$, the medium has been entirely invaded.
The dashed circle marks the location of an RGB edge belonging to the RGB cluster spanning the medium in the direction perpendicular to the paper plane, connecting the top and bottom sides of the cube in Fig.~2.
Inside the dotted circle, one sees the emergence of a loop disconnected from the spanning RGB cluster.
There, the surface dividing the R from the B fluid closely approaches the G-R division surface.
This leads to the presence of an RGB edge at the position indicated by the lower arrow.
Further above, where the G and B parts are in contact, the RGB network vanishes.
Finally, the upper arrow indicates where the R-B division surface reappears and another RGB edge emerges.
Given now the full, three-dimensional medium, not only a slice of it but the entire R part is connected and a loop emerges in the RGB network.
The sketch in Fig.~4 suggests a continuum picture, but on the lattice discretization effects are possible.
Partially developed loops appear as loops with a single edge.
These effects lead to spurious dangling ends attached to the backbone of the largest cluster.
Dangling ends are not present in the continuum where, in addition to the largest cluster, only disconnected loops are expected.
\section{Distance of disconnected loops from spanning cluster}
Figure \ref{fig::LoopDistanceDistribution} shows the number of loops separated by a distance larger or equal to $k$ from the spanning cluster as a function of ${k/L}$, where $L$ is the lattice length.
The distance of a disconnected loop to the spanning cluster is defined as the minimum distance between the loop and the spanning cluster in the dual lattice.
\begin{acknowledgments}
We acknowledge financial support from the ETH Risk Center and the Brazilian institute INCT-SC.
\end{acknowledgments}
\bibliography{rgb}
\end{document}